\documentclass[man]{apa7}
\usepackage[american]{babel}
\usepackage{csquotes}
\usepackage{amsmath, amssymb, mathtools}
\usepackage{graphicx}
\usepackage{etoolbox}
\usepackage{booktabs}
\usepackage{caption}
\usepackage{siunitx}
\usepackage{xfrac}
\usepackage{newtxtext}
\usepackage{newtxmath}
\usepackage[style=apa,backend=biber,sortcites=true]{biblatex}
\DeclareLanguageMapping{american}{american-apa}
\AtBeginBibliography{\interlinepenalty=10000}
\title{The "Days of Learning" Metric for Educational Evaluation }

\keywords{days of learning, effect size, education outcomes, charter school evaluation, national charter school study}

\shorttitle{Days of Learning}
\author{Gregory Camilli}
\authorsaffiliations{Rutgers University}

\authornote {Please send correspondence to greg.camilli@gse.rutgers.edu.}

\abstract{The third National Charter School Study (NCSS III) aimed to test whether charter school were effective and to highlight outcomes on academic progress. The authors reported that typical charter school students outperformed similar students in non-charter public schools by 6 days in mathematics and 16 days in reading. This "days of learning" metric used to claim relatively higher performance in charter schools than in comparable public schools. This logic of this metric is critiqued in this paper, and an alternative method of reporting outcomes is proposed.}

\begin{document}
\maketitle

\section {Introduction}
The third National Charter School Study (NCSS III) aimed to “test fundamental questions about charter school effectiveness and highlight outcomes and trends based on academic progress” \parencite{ray1}. The data for this study were drawn from 7,288 charter schools in 31 states, as well as the District of Columbia and New York City. Altogether over over 6.5 million student-year observations were collected with a matched sample of more than 1.85 million charter school students in traditional public schools.  The combined dataset represented approximately 81\% of tested public school students nationwide, making it one of the largest student-level datasets assembled for research on charter school performance. Given this comprehensiveness, the study has been extremely influential.

The results showed that typical charter school students outperformed similar students in non-charter public schools by 6 days in mathematics and 16 days in reading. The key metric used to claim better performance in charter schools is "days of learning." In this report, the origin of the days-of-learning metric is reconstructed with NAEP data, and the alternative interpretation of days of learning as within-year growth is considered.
\section{Background}
NCSS III uses the equivalence that .01 SD (standard deviation) effect size $\equiv 5.78$  days of learning. However, details on how this equivalence is calculated are limited. In response to criticism that the "days of learning" metric lacks a sufficient technical description \parencite{credo}, readers were directed to \parencite{han}, but "days of learning" is not mentioned in that document. Some additional clarification is available; however, technical details are still sparse:
\begin{quote}
\begin{description}
\item [A]  \, To make these results more meaningful to non-technical readers, we also include a transformation of the results in days of learning. As with standard deviations, the days of learning metric is expressed relative to the academic gain of the comparison student in a given year \parencite{ray2}.
\item [B] \, Using nationwide growth data from the National Assessment of Education Progress, the transformation involves multiplying the standard deviation units produced by our statistical analyses by 578 days. This yields 5.78 days of learning for every 0.01 standard deviation [SD] difference in our analysis\parencite{ray1}.
\end{description}
\end{quote}
It has also been proposed that "students on average gain about one third of a standard deviation per school year" with the important acknowledgment "that this correspondence has not been extensively researched and is likely to vary by grade level, position in the test distribution, and other factors" \parencite{han2}. Recently, this topic has received a good deal of scrutiny in the research literature. It is fairly well established, for instance, that within-grade achievement growth in most academic subjects decelerates over grade, accompanied by shrinkage of variability \parencite{stud} .Generally speaking, standardized school-year growth (fall-spring) is about 1 SD until grade 3, and thereafter, shows rapid deceleration. Moreover, growth also appears to depend on a starting benchmark such as the median of the test-score distribution \parencite{nwea}.This implies that .01 SD may a different interpretation across grades and students.

\section{Derivation}
The equivalence .01 SD effect size $\equiv$ 5.78  days can be closely approximated using sample estimates from the National Assessment of Education Progress (NAEP). For example, 2017 national public means and standard deviations (SD) of the scales for the grades 4 and 8 mathematics composite are 239.16 (31.7) and 281.96 (38.93), respectively. Using the simple average of the standard deviations, the standardized change from grade 4 to 8 can be calculated as 1.21 SD: 

\begin{equation}
1.21 =  \frac{281.96-239.16} {\sqrt{(31.70^2+38.93^2)/2}}
\label{eq:one}
\end{equation}

\vspace{0.5cm}
Next, 180 days in the typical school year result in 720 school days from grade 4 to grade 8, and the goal is to extrapolate how many days of 720 total are equivalent to .01 SD. In this case, the computation results in .01 SD $\equiv$ 5.94 days of learning:

\begin{equation}
5.94= \frac{(720 * .01)}{1.21}
\label{eq:two}
\end{equation}

This is close to the conversion factor used in NCSS III. Likewise, the corresponding means and standard deviations for the 2017 reading composite for grades 4 and 8 are 220.81 (38.08) and 265.33 (35.66). A similar procedure as above results in.01 SD effect size $\equiv$ 5.97. Although these calculations do not exactly reproduce the reported 5.78 value, the results suggest the days of learning equivalence based on the standardized difference between NAEP national averages and SDs for grades 4 and 8 (or possible 4-8 and 8-12 differences for some combination of NAEP assessments). 
\section {Comments}
There are three problems with this procedure. First, the grades 4 and 8 NAEP composite scores are not considered to have equivalent scales. As noted by the National Assessment Governing Board:
\begin{quote}
 NAEP scores cannot be calculated to figure out “grade level,” which depends on performance on local curriculum and tests. Thus, no increase or decrease in points on the NAEP scale may be equated to a jump in grade level \parencite{nagb}.
\end{quote}
This is one of the reasons why the field of educational measurement has moved away from “grade equivalent scores” \parencite{kor,this}. Second, the 4-year gap between grades 4 and 8 confuses within-grade and between-grade growth. Quote [A] above appears to support using within-grade growth. For example, in 2015, Oak Park Schools reported a difference of 0.89 SD in the district averages from the start (fall) and end (spring) of grade 4 \parencite{oak}.
If .89 SD corresponds to 180 school days, then 0.01 SD $\equiv$ 2 days of learning. Based on this calculation, the learning estimates in NCSS III are overstated by a factor of about $3$. Consequently, 6 days of learning in mathematics becomes 2 days, and 16 days in reading becomes 6 days. These positive "days of learning" values for within-grade growth still reflect a benefit to students, but it is clear that in some school districts, the reporting metric can greatly influence how charter school effectiveness is perceived. The third problem is that growth across grades is typically not linear, but rather has periodic ups and downs due to summer setbacks \parencite{allin}. This sawtooth pattern interferes with thinking of days of learning across assessment years as a "gain of the comparison student in a given year" (see quote A above).

There are related issues with the outcome variable in NCSS III, which is $z_2 - z_1$ or the difference in $z$-scores from period one (say grade 5) to period 2 (say grade 6). First, standardization does not create an equivalence between test content at different grade levels. On the surface the metric looks equivalent, but one could be comparing proficiency in basic number operations in period 1 to proficiency in decimals and fractions in year 2. Aggregating $z_2 - z_1$ across dozens of tests and grades further dilutes the interpretability of the days-of-learning metric. However, as in any summative evaluation, positive is still better than negative.

\section{Key Takeaways}
\begin{itemize}
\item In educational evaluations, quantitative results are often crafted into summary statistics and graphics to make them easier to communicate. However, alternative metrics may have very different implications for practice.
\item The "within-grade days of learning" metric appeals to practical experience with students’ growth within a school year. This enhances interpretability to a greater degree than the "between-grade days of learning" metric. Even so, the conversion may need to be targeted to a particular grade for maximum application.
\end{itemize}

For most audiences, days of learning should be presented along with contextualized effect sizes and clear explanations of the strengths and limitations of each metric. Within-grade days of learning is a critical competitor to between-grade days of learning \parencite{stuff}.

\printbibliography

@techreport{ray1,
  author = {Raymond, M. E. and Woodworth, J. L. and Lee, W. F. and Bachofer, S.},
  title = {National charter school study 2023},
  institution = {Center for Research on Education Outcomes, Stanford University},
  year = {2023},
  url = {https://credo.stanford.edu},
}

@misc{credo,
  author = {CREDO},
  title = {CREDO Response to Maul and Gabor},
  year = {2015},
  url = {https://credo.stanford.edu/wp-content/uploads/2021/09/CREDOResponsetoMaulandGabor3_000.pdf},
}

@article{han,
  author = {Hanushek, E.A. and Peterson, P.E. and Woessmann, L.},
  title = {Is the U.S. Catching Up? International and state trends in student achievement},
  journal = {Education Next},
  volume = {12},
  number = {4},
  pages = {24--33},
  year = {2012},
}

@techreport{ray2,
  author = {Raymond, M. E. and Woodworth, J. L. and Lee, W. F. and Bachofer, S.},
  title = {National charter school study 2013},
  institution = {Center for Research on Education Outcomes, Stanford University},
  year = {2013},
  url = {https://credo.stanford.edu},
}

@misc{nagb,
  author = {{National Assessment Governing Board}},
  title = {{A closer look at NAEP}},
  url = {https://www.nagb.gov/content/dam/nagb/en/documents/a-closer-look-at-naep.pdf},
}

@book{kor,
  author = {Koretz, D. M.},
  title = {Measuring Up: What Educational Testing Really Tells Us},
  publisher = {Harvard University Press},
  year = {2008},
}

@article{this,
  author = {Thissen, D.},
  title = {Validity issues involved in cross-grade statements about NAEP results},
  journal = {Educational Measurement: Issues and Practice},
  volume = {31},
  number = {1},
  pages = {3--13},
  year = {2012},
  doi = {10.1111/j.1745-3992.2012.00233.x},
}

@misc{oak,
  author = {{Oak Park Schools}},
  title = {2015 Mathematics Student Status Norms},
  year = {2015},
  url = {https://www.oakparkschools.org/parents/nwea-map-assessment/normative-data-charts/#2015math},
}

@Inbook{stuff,
author="Stufflebeam, Daniel L.
and Shinkfield, Anthony J.",
title="Scriven's Consumer-Oriented Approach to Evaluation",
bookTitle="Systematic Evaluation: A Self-Instructional Guide to Theory and Practice",
year="1985",
publisher="Springer Netherlands",
pages="311--342",
doi="10.1007/978-94-009-5656-8_10",
}

@article{stud,
  author    = {Student, S. R.},
  title     = {Vertical Scales, Deceleration, and Empirical Benchmarks for Growth},
  journal   = {Educational Researcher},
  volume    = {51},
  number    = {8},
  pages     = {536--543},
  year      = {2022},
  doi       = {10.3102/0013189X221105873},
}

@techreport{han2,
  author    = {Hanushek, E. and Woessmann, L.},
  title     = {The economic impacts of learning losses},
  institution = {OECD Publishing},
  year      = {2020},
  number    = {225},
  series    = {OECD Education Working Papers},
  address   = {Paris},
  doi       = {10.1787/21908d74-en},
  url       = {https://doi.org/10.1787/21908d74-en},
}

@techreport{nwea,
  author    = {NWEA},
  title     = {Growth and Norms: 2020 MAP Normative Data Overview},
  institution = {NWEA},
  year      = {2020},
  url       = {https://teach.mapnwea.org/impl/maphelp/Content/Data/GrowthInsights.htm#Norms},
}

@article{allin,
  author = {Allington, R. L. and McGill-Franzen, A.},
  title = {The Impact of Summer Setback on the Reading Achievement Gap},
  journal = {Phi Delta Kappan},
  volume = {85},
  number = {1},
  pages = {68--75},
  year = {2003},
  doi = {10.1177/003172170308500119},
  note = {Original work published 2003},
}

\end{document}